\documentclass{emulateapj}
\usepackage{apjfonts}

\slugcomment{accepted for publication in ApJ}
\begin{document}

\def\arcdeg{\hbox{$^\circ$}}
\def\arcmin{\hbox{$^\prime$}}
\def\arcsec{\hbox{$^{\prime\prime}$}}
\def\lessim{\mathrel{\hbox{\rlap{\hbox{\lower4pt\hbox{$\sim$}}}\hbox{$<$}}}}
\def\gtrsim{\mathrel{\hbox{\rlap{\hbox{\lower4pt\hbox{$\sim$}}}\hbox{$>$}}}}
\newcommand{\kp}{$K^\prime$}

\title{The Submillimeter Properties of Extremely Red Objects in the CUDSS  Fields}

\author{T.M.A Webb\altaffilmark{1,2},  M. Brodwin\altaffilmark{2,4}, S. Eales\altaffilmark{3}, S.J. Lilly\altaffilmark{5}}

\altaffiltext{1}{Leiden Observatory, Niels Bohrweg 2, 2333 CA Leiden, The Netherlands}
\altaffiltext{2}{Visiting Astronomer, Canada-France-Hawaii Telescope, Operated by the National Research Council of Canada, the Centre de la Recherche Scientifique de France, and the University of Hawaii}
\altaffiltext{4}{Department of Astronomy and Astrophysics, University of Toronto, 60 St George St, Toronto, 
Ontario, Canada, M5S 1A1}
\altaffiltext{3}{Department of Physics and Astronomy, Cardiff University, P.O. Box 913, Cardiff, CF2 3YB, UK} 

\altaffiltext{5}{Department of Physics, Swiss Federal Institute of Technology (ETH-Zurich), ETH Hoenggerberg, CH-8093, Zurich, Switzerland}

\begin{abstract}
We discuss the submillimeter properties of Extremely Red Objects (EROs) in the two   Canada-UK Deep Submillimeter Survey (CUDSS) fields. We measure the mean submillimeter flux of the ERO population (to {\kp}$<$ 20.7) and find  0.40 $\pm$ 0.07 mJy for EROs selected by $(I-K)>$ 4.0  and 0.56 $\pm$ 0.09 mJy for EROs selected by  $(R-K)> $ 5.3 but,  these measurements are dominated by discrete, bright submillimeter sources. We estimate that EROs produce  7-11\% of the far-infrared background at 850\micron. This is substantially less than a previous measurement by \citet{weh02} and we discuss possible reasons for this discrepancy.    We show that ERO counterparts to bright submillimeter sources lie within the starburst region of the  near-infrared color-color plot of  \citet{poz00}.  Finally, we claim that pairs or small groups of EROs with separations of $\lesssim$ 10 arcsec often mark regions of strong submillimeter flux. 

\end{abstract}


\section{Introduction}

A complete accounting of all the systems which produce the extragalactic background light is key to our understanding of galaxy assembly and evolution. Observations of the early universe  have revealed a  wide variety of objects  whose relationship to each other,  importance to galaxy formation, and  contribution to the extragalactic background energy remains unclear. The fact that these systems have been selected over different wavelength ranges and using different selection criteria has made it is difficult to determine if we are studying different aspects of the same objects, or completely orthogonal populations.  Many of these objects are found over similar redshift ranges and are co-evolving, while others may be linked through time by evolutionary processes.    

At near-infrared wavelengths a population of red objects begins to emerge at faint, $K\gtrsim$ 18, magnitudes \citep{els88,mcc92,hu94}. This class of systems,  labeled Extremely Red Objects (EROs), is photometrically defined by NIR color, typically $(I-K)>$ 4 or  $(R-K)> $  5 $-$ 5.3.  Two different galaxy populations are selected  by this color definition: an old stellar population at $z\sim$1, or alternatively, a young starburst at high-redshift, deeply enshrouded in dust.  Low-mass stars and brown-dwarfs within the Galaxy also have similar NIR colors but have such low surface densities relative to faint galaxies that they do not significantly contaminate ERO samples.  

Determining the fraction EROs belonging to each category has been difficult work, hindered by the faintness of these objects at optical and NIR wavelengths. Spectroscopic observations indicate that  evolved stellar populations and dusty starbursts are present in roughly equal numbers \citep{dun96,gra96,coh99,cim02a} and this picture is generally supported by larger studies employing  less expensive, but more ambiguous,  techniques such as color \citep{poz00,mar01,sma02a}, radio emission \citep{sma02a}, or morphology \citep{yan03,gil04}. There is substantial variation between the results however, stemming from different survey depths, varying color-selection criteria, and clustering effects \citep{dad00,mccar01,smi02,sma02a,dad02,yan03}.

The detection of submillimeter emission from an ERO is a direct confirmation of a starburst nature (ignoring for the moment the problem of confusion within the beam). At current observational limits however submillimeter imaging can only detect the most luminous star forming EROs, those that overlap with the discrete submillimeter-bright population  at high-redshift  discovered with SCUBA (SMGs)  \citep{sma97, bar98, hug98, eal00, sco02} which  are sites of  perhaps the most substantial star formation in the high-redshift universe.  Though still suffering from small number statistics $\sim$20-30\% of the systems detected deep submillimeter surveys  are EROs (when observed in the NIR)  \citep{sma99,ivi02,cha03a,web03c} though this fraction obviously depends on the relative depths of the the 850{\micron} and $K$-band imaging and the uncertainty in the in the identifications of the counterparts of non-radio detected SMGs.  What is not clear, however, is how many of the $K$-selected EROs are discrete, bright submillimeter sources and the submillimeter propeties of  EROs below the current detection limits of submillimeter surveys. These less luminous systems, though not individually submillimeter-bright could together produce a substantial fraction of the FIR background light. Two previous studies have attempted to address  these issues through statistical studies EROs \citep{moh02,weh02} and the results of both groups indicate that as a population these systems  emit substantial submillimeter flux. In particular, \citet{weh02} (hereafter W02) found $K^{\prime} <$ 21.6 EROs produce $\sim$50\% of the FIR background light at 850$\mu$m, with  this contribution peaking at $K^{\prime}\sim$20.    However, as with all ERO studies, these results may depend on  ERO selection criteria and field-to-field variance.

Thus, the relationship between EROs and bright SMGs, and the submillimeter properties of  EROs below the detection limit of SCUBA, remain  open questions, as does the ERO  contribution to the far-infrared background energy. Those EROs which are extremely luminous in the submillimeter are a minority of both the ERO population and the SMGs (at least to current $K$ depths) but must represent an important phase of galaxy evolution. Understanding their natures will provide a key link between these two populations. Isolating  the properties of EROs which are individually or statistically submillimeter bright which  differentiate them from the rest of the ERO population is an important step toward this understanding.  Such properties may include  the NIR colors and magnitudes \citep{poz00,weh02},  morphologies, redshifts and the local environment.  Environment plays a key role in galaxy evolution and there is growing evidence of a correlation between submillimeter emission and high-density regions or galaxy merging and interactions \citep{ivi00,sma03,ale03,web03c,ner03,cha03c} and this may be  be an important driver of star-formation in the ERO population. Additionally, the different redshift distributions of the EROs and SMGs (as far as we know them) \citep{cha03b} hint at an evolutionary relationship between these populations perhaps with SMGs fading into less luminous star-forming EROs at later times.  Addressing these issues will not only illuminate the natures of these two populations but the process of galaxy evolution in general.

The paper is organized as follows. In \S2 and \S3  we present the data and define our ERO samples. In \S 4 we measure the measure  submillimeter flux of EROs. We discuss the ERO contribution to the extragalactic background light in \S 5. In \S 6  we present the \citet{poz00} color-color plot for these systems. In \S 7 we  discuss the effect of ERO companions in the analysis. We finish in \S 8 with some thoughts difference between submillimeter-bright and faint EROs.  Throughout we have used a flat, $\Lambda$=0.7 cosmology and H$_{\circ}=$ 72 km/s/Mpc.

\section{Observations}

\subsection{Submillimeter Data}

The Canada-UK Deep Submillimeter Survey (CUDSS) was a deep, random-field submillimeter survey carried out from 1998-2002 at 850{\micron} using SCUBA on the JCMT. The submillimeter data and catalogs are discussed extensively in \citet{eal00} and \citet{web03b}.  The survey consists of two fields, CUDSS14 and CUDSS03, which cover 48 arcmin$^2$ and 60 arcmin$^2$ respectively,  to a roughly uniform  3$\sigma$ depth of 3 mJy.  The  chopping and nodding technique employed by SCUBA to remove sky-flux creates distinct beam pattern on the maps: each positive source is flanked by two negative ``sources''  with half the flux-density, offset from the positive source by a distance equal to the chop throw (in our survey this is 30\arcsec).  This results in a map which has a total flux of zero.   In addition to the bright source catalog (i.e. $S_{850{\mu}m} >$ 3 mJy) the contiguous nature of the maps allow an  extraction information to  fainter flux levels \citep{web03a}. 

\subsection{Optical and NIR Imaging}

The two CUDSS fields were originally selected to lie within Canada-France Redshift Survey fields \citep{lil95}  and are well studied at many wavelengths. The Canada-France Deep Fields Survey has obtained new deep optical imaging, including $RIz$, over these areas \citep{mcc01,bro04} using the CFHT 12K camera.  The pixel scale of these data is  0.207 arcsec/pixel, and reaches 2$\sigma$ depths of $I$=25.58 (25.06), $R$=25.72 (25.72), and $z$=24.55 (24.75) for CUDSS03 (CUDSS14).   

The {\kp} data were obtained in 2000 with the CFHT-IR camera on the Canada-France-Hawaii Telescope (CFHT) \citep{hut02,web03c}.   The data cover  $\sim$ 2/3 of the two CUDSS fields, and areas of 35.4{\arcmin} and 55.8{\arcmin} for CUDSS14 and CUDSS03 respectively.  The CFHT-IR camera field-of-view is 3.6\arcmin$\times$3.6{\arcmin} and the final images are mosaics of many individual pointings.  The pixel scale is identical to that of the $RIz$ data, 0.207 arcseconds/pixel, and conditions during the run were stable with seeing typically 0.8\arcsec. The depth over the two fields is not uniform but a  5$\sigma$ limit of  $K^{\prime}$= 20.7 is reached over most of the area.   

\section{ERO Sample Selection}
\begin{figure}
\plotone{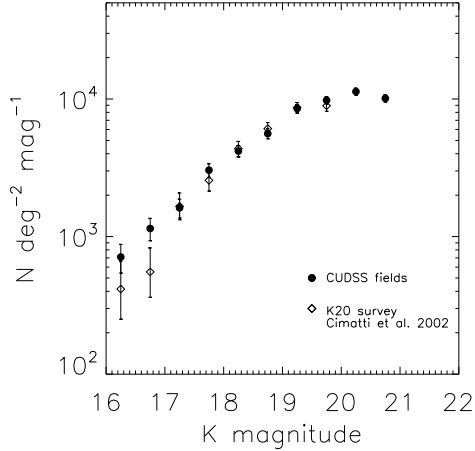}
\caption{The K number counts for both fields, CUDSS14 and CUDSS03. Shown for comparison are the number counts from the K20 Survey. \citep{cim02b}\label{kcounts}}
\end{figure}

Object detection was performed on  the {\kp} images using the source extraction package SExtractor \citep{ber96} and magnitudes were extracted  through 3{\arcsec} aperture photometry. The {\kp} source counts are presented in Fig.\ref{kcounts}  and are in good agreement with other surveys, though for clarity we plot only the K20 counts \citep{cim02b}.  Colors were measured in 2.5{\arcsec} apertures on images degraded to common seeing of 1.3{\arcsec}. The effects of  galactic extinction were removed in both fields \citep{mcc01}

\begin{figure}
\plotone{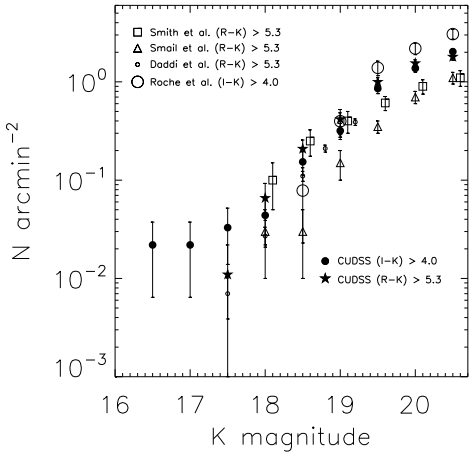}
\caption{The number counts for EROs selected by $(I-K)>$ 4.0 and $(R-K)>$  5.3  for both fields, CUDSS14 and CUDSS03. Shown for comparison are the number counts from \citet{roc03}, \citet{smi02}, \citep{dad00}, and \citet{sma02a} (see legend).  \label{erocounts}}
\end{figure}

We selected EROs by two criteria commonly used in the literature, $(I-K)>$ 4 and $(R-K)>$ 5.3.  For objects without $R$ or $I$ detections or detections at $<$ 2$\sigma$ (as determined from the average noise properties of the $R$ and $I$ images) the $2\sigma$ limit on their magnitude was adopted.  This provides a lower (blue) limit  on their color, and they were taken to be EROs if this limit met the above selection criteria.  We present the ERO source counts (as a function of {\kp} magnitude) in Fig.~\ref{erocounts} and compare them with other surveys which apply the same selection criteria. We find the source counts of each field, using both color criteria, are in good agreement with these other surveys and therefore conclude that these fields do not contain exceptional over or under-densities of EROs.

In CUDSS03 (CUDSS14) we select 130 (111)  objects with $(I-K)>$ 4 and 77 (87) with $(R-K)>$ 5.3.    There are 51 and 77 galaxies in common between the two samples for CUDSS03 and CUDSS14 respectively.  Thus,  $(I-K)>$ 4.0  appears to be a more generous ERO definition than $(R-K)>$ 5.3, an effect also seen by  \citet{sma02a} . However. the  reddest colors we can measure for our faintest $K$ magnitudes are $(I-K)=$ 5.0 ($(I-K)=$ 4.6) for CUDSS03 (CUDSS14) and $(R-K)=$ 5.2 for both fields and therefore we probe EROs to slightly deeper K-magnitudes using $(I-K)$-selection.

\section{The submillimeter flux of EROs}
\subsection{The overlap of the ERO and discrete submillimeter populations} 

A number of recent papers have investigated the near-infrared properties of SMGs which have been robustly identified through radio detections \citep{ivi02,cha03a,web03c}.  While these systems show a range of near-infrared colors, and some are composite  systems with components of different colors, the average $(I-K)$ color for these systems is significantly redder than that of the background $K$-selected population, and many SMGs are classified as EROs.  

In the CUDSS catalog there are ten  bright (discrete) SMGs ($S_{850{\micron}} > $ 3mJy)  whose most likely counterpart is an ERO (Table \ref{brightsmg}) and these objects are discussed in detail in  \citet{web03c} and \citet{cle03}. These papers also describe the statistical identification methods which were used to select these counterparts, but we briefly outline them again here. We  worked with radio (1.4 GHz) and $IK$ imaging and selected the statistically most likely counterparts based on positional probability analysis using surface densities determined from either the 1.4 GHz data or $I$ and $K$ magnitudes and colors. In \citet{web03c} we showed that the $IK$ identification algorithm blindly and unambiguously selected the same counterpart as the radio data.  We concluded that if  the SMG source properties are similar to the radio-identified systems at redshifts beyond the reach of the radio, the $IK$ identification algorithm will remain successful at selecting the correct counterparts.  Three of the ten  EROs are  detected at 1.4GHz which allows us to state with further confidence that  these particular identifications are correct. The remaining seven, however, are not detected in the radio (perhaps because they lie at redshifts beyond the radio limit) and so, although statistically very likely, some (or all)  of these identifications could be incorrect.  Thus, between 6-20\% of the discrete SMGs in the CUDSS catalogue  have ERO counterparts to $K$ = 20.7.  

\begin{deluxetable*}{llllll}
\tabletypesize{\scriptsize}
\tablecaption{EROs with possible SCUBA detections at $>$ 3$\sigma$ \label{brightsmg}}
\tablewidth{0pt}
\tablehead{
\colhead{SCUBA ID} & \colhead{$K$}   & \colhead{$(I-K)$}  & \colhead{$(R-K)$} & \colhead{S$_{850{\mu}m}$} & \colhead{Note}
}

\startdata

14.1 &  19.3 $\pm$ 0.1 & 5.1 $\pm$ 0.3 & $>$ 6.3 & 8.7 $\pm$ 1.0 & 1.4GHz det.  \\
14.2 &  19.6 $\pm$ 0.1 & 4.9 $\pm$ 0.4 & 5.7 $\pm$ 0.4  & 5.5 $\pm$ 0.9 & ...   \\
14.5 &  20.7 $\pm$ 0.1 & 4.1 $\pm$ 0.5 & $>$ 4.9   & 4.6 $\pm$ 1.0 .& ... \\
14.7 &  20.6 $\pm$ 0.1 & 4.1 $\pm$ 0.5 & $>$ 4.9    & 3.2 $\pm$ 0.9 & ...  \\
14.8 &  19.7 $\pm$ 0.1 & 4.3 $\pm$ 0.3 & 4.8 $\pm$ 0.2 & 3.4 $\pm$ 0.9 & ... \\
14.9 &  19.2 $\pm$ 0.1 & $>$ 5.7   & $>$ 6.3  & 4.3 $\pm$ 1.0 & 1.4GHz det.\\
14.23 & 20.0 $\pm$ 0.1 & $>$ 4.7  & $>$ 5.3 & 2.8 $\pm$ 0.9 & ...\\
3.4 & 19.8 $\pm$ 0.1 & 4.0 $\pm$ 0.2 & $>$ 4.6 & 8.0 $\pm$ 1.0  & ... \\
3.7 & 20.4 $\pm$ 0.1 & $>$ 5.0  & $>$ 5.1   & 8.2 $\pm$ 1.5 & 1.4GHz det.\\
3.22 & 19.6 $\pm$ 0.1  & 4.2 $\pm$ 0.2  & 5.7 $\pm$ 0.5 & 3.1 $\pm$ 1.0 & ... \\
\enddata

\end{deluxetable*}

\subsection{The mean submillimeter flux of the  EROs }

To estimate the mean 850$\mu$m flux of the ERO population we measure the flux on the SCUBA map at the {\kp}-selected position of each ERO (see also \citet{web03a}).   The  combination of the large beam size of the JCMT at 850{\micron} (15{\arcsec} FWHM)  and the chopping technique (30{\arcsec} in EW)  means the measured flux of an ERO could be contaminated by a nearby bright submillimeter source which is unrelated to the ERO.  For example, if the location of an ERO fell within the negative off-position of a bright submillimeter source we  would measure an erroneous {\it negative} flux for the ERO. To reduce this effect we have removed all of the discrete submillimeter sources from the maps  which are securely identified (i.e. identified through radio-detections) with non-EROs  \citep{eal00,web03b,web03c,cle03}  before extracting the ERO fluxes.  We then compute  the noise-weighted mean of these measurements (see \citet{eal00,web03b} for noise discussion).  These results  are listed in Table \ref{smmflux}. In practice, removing the discrete submillimeter sources which are securely identified with non-EROs has not significantly (i.e.~within the uncertainties) changed the mean flux measurements. It has lead to an increase in the measured mean flux of 0.12mJy for CUDSS03 and a decrease of 0.05mJy for CUDSS14 (remembering the change can be either negative or positive because of the negative-positive nature of the beam profile) and thus the combined measurement has remained constant within the uncertainties.

Though we use the true {\kp}-determined position for the EROs, the submillimeter-determined position (if it were possible to measure) will always be offset from this location due to noise and confusion \citep{eal00,web03c}.  At the {\kp}-position there is an equal chance of noise boosting the signal up or down, while if we attempt to search for the peak flux of th ERO we would be pulled toward positive noise spikes and thus the measurement of the average submillimeter flux would be biased upward.   To investigate the effect of the beam on the mean measurement we have performed this analysis for the 850{\micron}-detected radio sources in the CUDSS14 field (i.e.~we measure the average submillimeter flux at the radio-determined position).  The true average submillimeter flux of these systems is 5.4 $\pm$ 0.4 mJy and our recovered average is 5.2 $\pm$ 0.4 mJy. Thus this technique recovers the true average flux within the uncertainties.     

There are two further complications  in this analysis, that we have not yet considered.  (1) Are the results strongly affected by confusion? (2) To what degree are these measurements dominated by a small number of very bright submillimeter sources? We have already addressed the issue of confusion between the EROs and those  submillimeter-sources which are {\it not} EROs  by removing their submillimeter flux from the map but  we have not discussed  confusion between EROs themselves.  If two or more EROs are separated by a distance roughly less than the  beam size ($\sim$ 15{\arcsec}) they will be confused and their measured flux densities will be incorrect. In the simplest case where two EROs lie directly on top of one another, each emitting 1mJy at 850$\mu$m, we would erroneously measure 2 mJy for each.  As we will discuss in more detail in \S 7 a number of the EROs associated with discrete SMGs  have apparent  ERO companions (\S 4) and therefore the n\"aive treatment outlined above will lead to an over-estimate of the mean submillimeter flux. 

We now employ a more sophisticated statistical analysis,   accounting for this confused flux in the following way.  Consider two confused EROs with true submillimeter fluxes of $I_1$ and $I_2$ and separated by a distance $r$ which is less than the approximate beam size.  The measured fluxes $f_1$ and $f_2$ may be written as follows: 

\begin{equation}
f_1 = I_1 + I_2 e^{(-r^2/2{\sigma})}
\end{equation}
\begin{equation}
f_2 = I_2 + I_1 e^{(-r^2/2{\sigma})}
\end{equation}  

As we know $f_1$, $f_2$, and $r$ we may  solve for $I_1$ and $I_2$.  These equations are easily expanded for any number of confused objects though in these data groups of more than three EROs are rare and  there are no groups which contain more than four objects.  Using this procedure will result in an improved estimate of the  mean submillimeter flux for the ERO sample as a whole though the measured flux densities of individual EROs may be incorrect since in reality only one of the confused EROs may be responsible for the submillimeter emission. 

We present the average submillimeter flux for the different samples, obtained using this method, in Table \ref{smmflux}.    The beam correction is important for 14\% (11\%) of the EROs in  the $(I-K)>$ 4 ($(R-K)>$ 5.3) sample, and the corrections to their individual flux measurements vary from $\sim$ 10 - 50\%.  For example, source CUDSS03.7 (shown in Figure 6) contains three EROs within 4{\arcsec} of eachother, and within the beam of a discrete SMG. The  flux measurements for these three EROs before the beam correction are 7.7 mJy, 7.3 mJy, and 6.6 mJy, with a non-weighted average of 7.2 mJy. After the beam correction has been applied these fluxes become 6.2 mJy, 3.8 mJy and 3.8 mJy respectively, with a non-weighted average of 4.6 mJy. Looking at  the entire sample  this method has lead to a decrease in the measured mean flux of $\sim$ 20\% (Table \ref{smmflux}).

\begin{figure}
\epsscale{2.0}
\plottwo{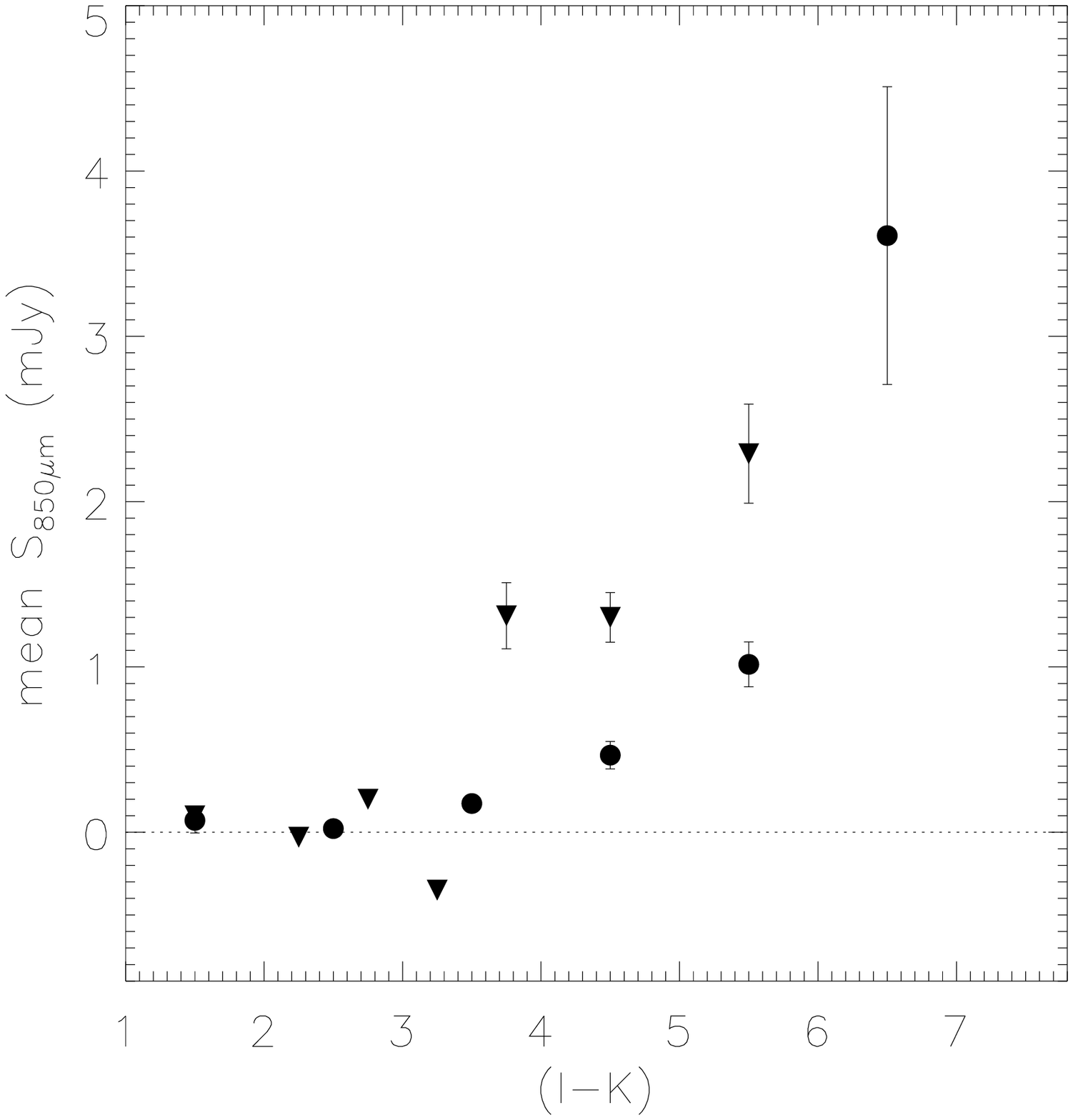}{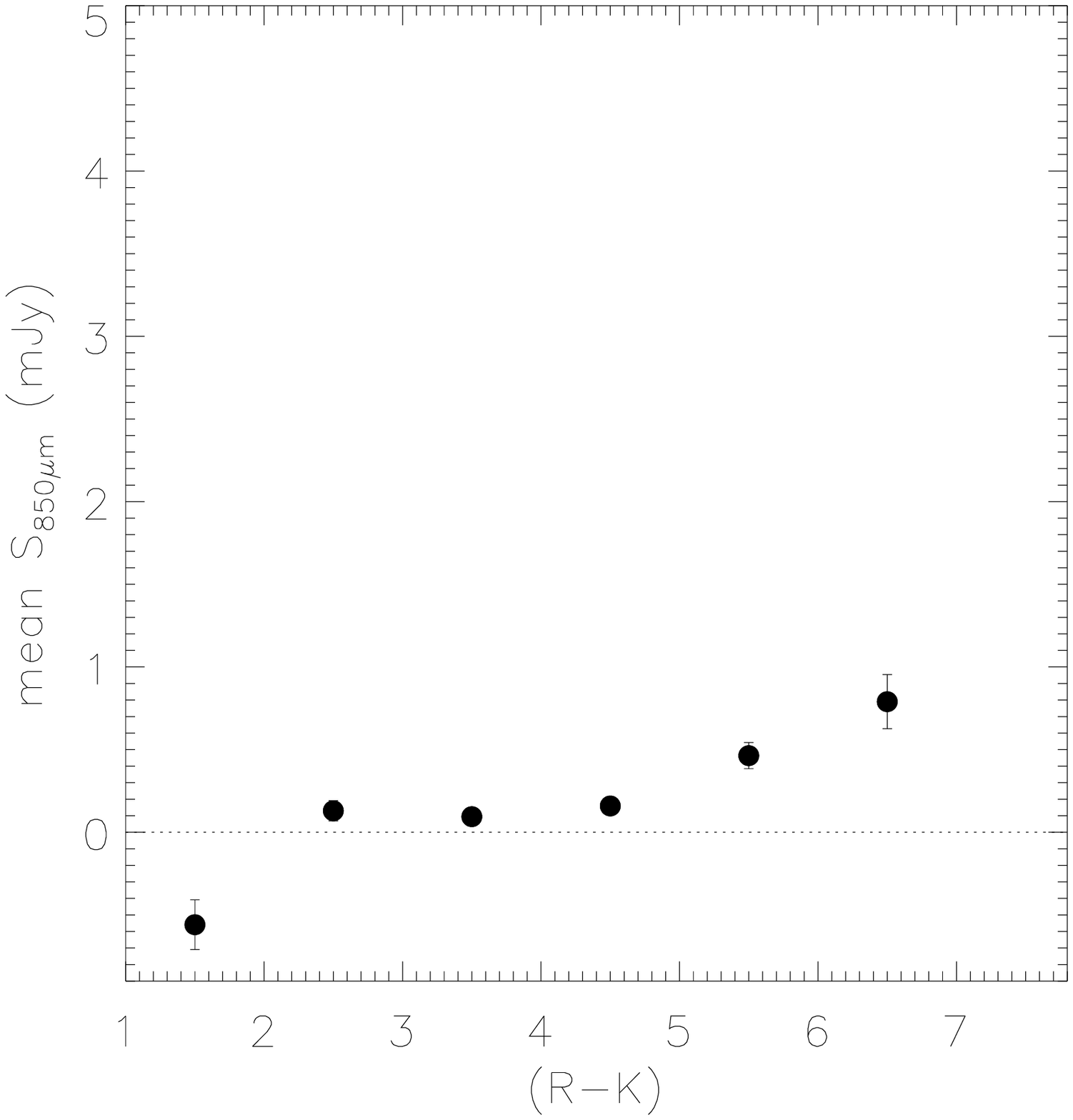}
\caption{  The mean 850{\micron} flux measured for EROs as a function of their NIR color. Top: The $(I-K)$ color. The solid circles correspond to this work and the solid triangles correspond to \citet{weh02}. Bottom: The $(R-K)$ color.   \label{colsmmp}}
\end{figure}

In Fig.~\ref{colsmmp} we show  the mean 850{\micron} flux the K-selected galaxies in both fields as a function of their color, for both color definitions. Clearly, the mean submillimeter flux is a strong function of color, with redder objects producing more submillimeter emission, regardless of the color definition.  We also plot the results of W02, denoted by the filled triangles and  same trend is seen, though W02 measure systematically higher flux above $(I-K)\gtrsim$ 3.5.

The question of whether  these measurements are dominated by a small number of  bright objects remains.  W02 concluded that their signal did not arise from a minority of objects; after removing  the five most positive and five most negative measurements  from their analysis they saw little change in the measured value.  If we perform a similar analysis on these data our detection also remains significant. This does not mean, however, that the signal is not dominated by a minority of sources.  Firstly, out of the few hundred EROs in our sample, ten are the best identified counterparts of SMGs and simply removing five of these does not fully address the issue of all ten dominating the emission.  A more important point, however, which we have already touched on and which  we discuss  later in \S 7, is that these submillimeter-bright EROs are often found in pairs or triplets with  other EROs. Thus, though ten ERO (defined by $(I-K)>$ 4) have been identified as SMG counterparts, 17 (again for $(I-K)>$ 4) are associated with discrete SMGs. By associated we mean that they are found within the beam of a discrete SMG. The mean measurement is dominated by these EROs, which themselves are a minority of the ERO population and which are associated with an even smaller number of discrete SMGs.  As we show in Table \ref{smmflux}, significant flux is not detected from EROs which are not found within the confused beam  of a bright, discrete SMG.

\begin{deluxetable*}{lllll}
\tabletypesize{\scriptsize}
\tablecaption{The Submillimeter Flux of EROs \label{smmflux}}
\tablewidth{0pt}
\tablehead{
\colhead{Field} & \colhead{ERO criteria}   & \colhead{(simple) mean}  & \colhead{with beam } & \colhead{``bright'' EROs} \\
\colhead{} & \colhead{}  & \colhead{$S_{850{\mu}m}$ (mJy) } & \colhead{correction} & \colhead{removed\tablenotemark{a}} 
}

\startdata

both fields & $(I-K)>$ 4.0 & 0.59 $\pm$ 0.07 & 0.40 $\pm$ 0.07 & \phantom{-}0.10 $\pm$ 0.07 (17)  \\
both fields & $(R-K)>$ 5.3 & 0.61 $\pm$ 0.09 & 0.56 $\pm$ 0.09 & \phantom{-}0.22 $\pm$ 0.09 (11) \\ 
CUDSS+03 & $(I-K) >$ 4.0  & 0.36  $\pm$  0.11  & 0.31 $\pm$ 0.11 & \phantom{-}0.16 $\pm$ 0.12 (5) \\
CUDSS+03 & $(R-K)>$ 5.3  & 0.12 $\pm$  0.16  &  0.11 $\pm$ 0.16 & -0.05 $\pm$ 0.17 (2) \\
CUDSS+14 & $(I-K)>$ 4.0   & 0.74  $\pm$ 0.09  & 0.45 $\pm$ 0.09 & \phantom{-}0.06 $\pm$ 0.10 (12) \\
CUDSS+14 & $(R-K)>$ 5.3  & 0.82  $\pm$ 0.11 & 0.76 $\pm$ 0.11 & \phantom{-}0.34 $\pm$ 0.12 (9) \\

\enddata
\tablenotetext{a}{the number in brackets denotes the number of EROs removed from the analysis}
\end{deluxetable*}

\section{The contribution of EROs to the FIR background light}

Accounting for all of the objects which produce the cosmic background light is crucial for our understanding of the cosmic star formation history.  W02 estimate that EROs, with $(I-K)> $ 4.0 and  {\kp} = 21.6 produce $\sim$ 50\% of the integrated background at 850{\micron}, and that half of this energy is emitted by objects with {\kp}$<$ 20.   Taking our mean 850{\micron} flux, and the surface density of EROs in our fields we estimate that EROs produce 7-11\%  of the background at 850{\micron} where the range in values encorporates the two ERO selection techniques we have used and the values for the background given by \citet{pug96} and \citet{fix98}. Clearly a marked disagreement exists  between these results and W02. The fact that W02 reach one magnitude deeper in {\kp}cannot resolve this disagreement since they estimate that half of their measured flux is emitted from objects with {\kp}$<$ 20.

The difference in these two estimates stem from a larger surface density of EROs in the W02 fields and a higher mean 850{\micron} flux per ERO, and we discuss   each in turn.  The W02 data cover 11.5 arcmin$^2$ over three cluster fields and at {\kp} = 21.6 they find a surface density of 8 arcmin$^{-2}$ which is significantly higher than seen over larger areas \citep{smi02}. Though their source counts are dominated by a single field when this field is removed the surface density still remains high (though within acceptable range). If the surface density measured by \citep{smi02} is used this reduces their estimate of the background contribution by a factor of 0.7.   We have covered roughly ten times the area of W02 and therefore our surface density should be a better reflection of the true surface density of EROs.

The second issue is the mean 850{\micron} flux per individual ERO.  The largest difference here may stem from the effect of the beam correction (\S 4.2) which does not appear to have been included in the W02 analysis.  In the case of our data this correction  reduced the measured flux by roughly 20\%.  At our deepest limits we have a source density of $\sim$ 2  arcmin$^{-2}$ and,  assuming no clustering, this corresponds to an  average separation of EROs of roughly 48{\arcsec}, and the beam correction is important for $\sim$12\% of the EROs.  At the surface density of W02 sample of 8 arcmin$^{-2}$ the average separation of sources of about half this value (24{\arcsec}) and the beam correction will be important for 45\% of the EROs. If the highest density field of W02 is removed then about 19\% of the sources will be effected.  Thus, even for the lowest surface density used by W02 the correction  is likely  $\sim$ 30\% (and more than 50\% if the highest surface density is used).

 We can also make some very simple arguments against such a large ERO contribution to the background light.  We claim that the mean 850{\micron} flux of EROs is dominated by a minority of discrete SMGs with $S_{850{\micron}}> $ 3mJy.  In blank field submillimeter surveys such objects produce $\sim$15-30\% of the FIR background at 850{\micron} \citep{bla99,bar99,cow02,sma02b,web03b} and 20\% of these objects have  possible ERO  counterparts to our magnitude-limits (\S 4). Therefore, a contribution on order  6\%  from these EROs is expected. The lack of a significant detection of EROs below the 3 mJy means that the total contribution to the background cannot be much higher than this.
 
\begin{figure}
\epsscale{2.0}
\plottwo{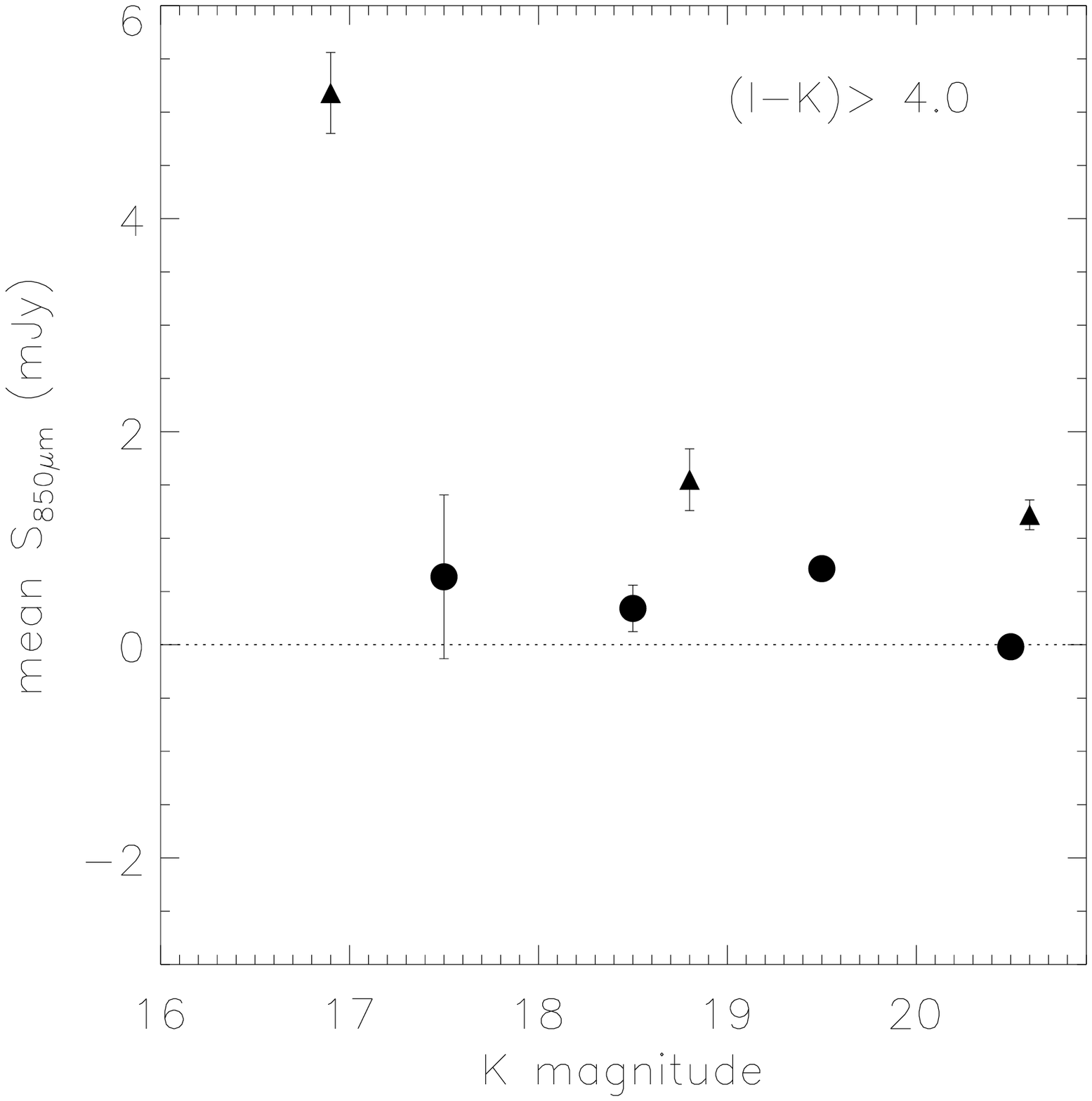}{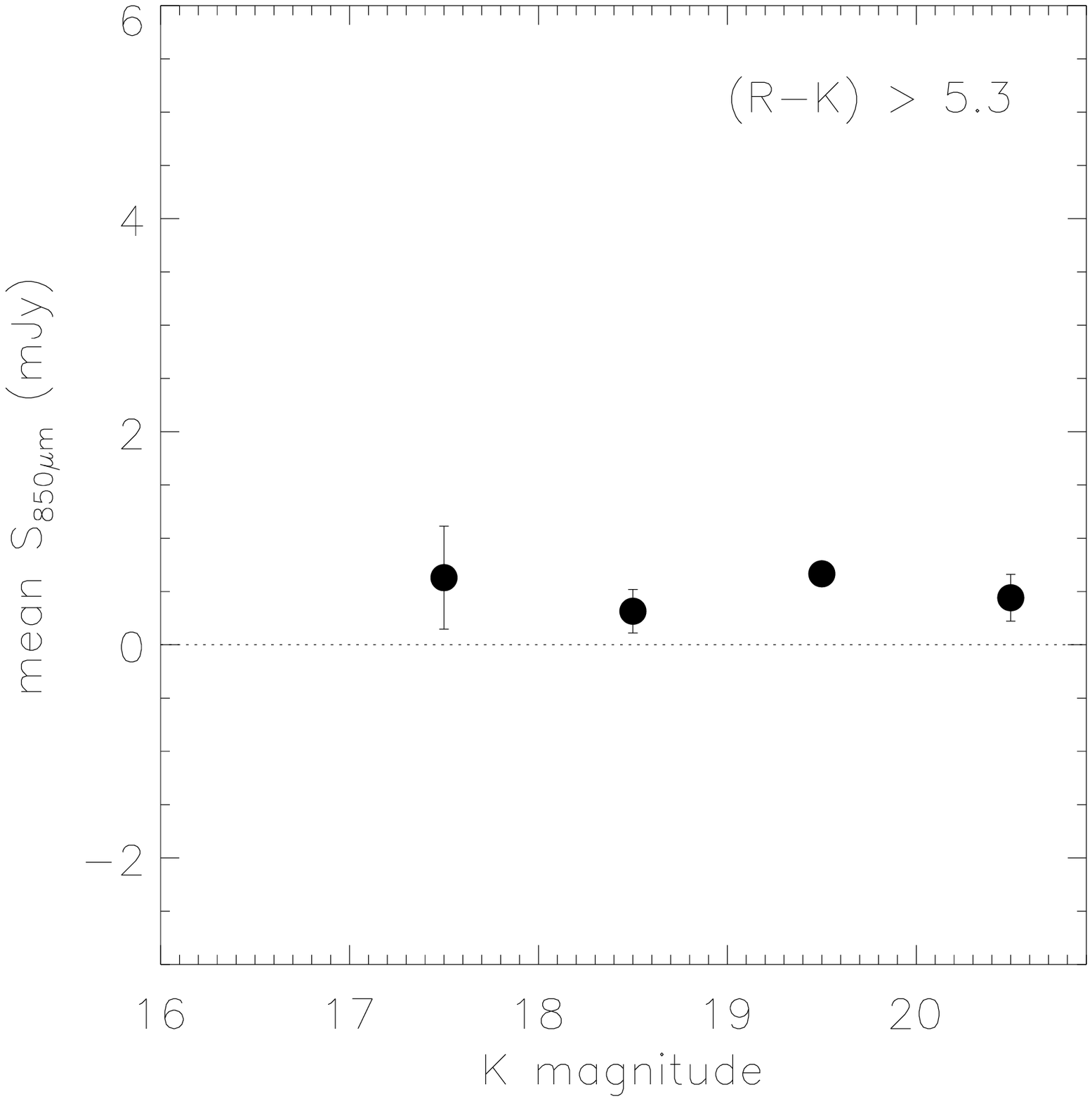}
\caption{ The mean 850{\micron} flux measured for EROs as a function of their $K$ magnitude. Top:  The $(I-K)$-selected sample. The solid circles correspond to this work and the solid triangles correspond to \citet{weh02}. Bottom: The $(R-K)$-selected sample.   \label{magsmmp}}
\end{figure}

In Fig.~\ref{magsmmp} we plot the mean 850{\micron} flux of EROs in our sample as a function of their K-magnitude, and over-plot the measurements of W02.   W02 reported a strong increase in submillimeter flux to brighter  magnitudes which is not reflected in our data. This is difficult to understand as it implies too many {\kp}-bright counterparts to discrete SMGs.  Specifically, the average 850{\micron} flux of {\kp}$\sim$ 19 EROs in the W02 sample is $\sim$ 3mJy and the integrated source counts of EROs at this magnitude roughly equals the integrated source counts of  discrete SMGs with $S_{850{\micron}}>$ 3mJy. This implies that a significant fraction of the discrete SMG population should have ERO counterparts brighter than {\kp}$\sim$ 19, and it is well established that this is not the case \citep{web03c,cle03,sma02b,ivi02}. This holds even if we take the more realistic view that the average flux measurement reflects a wide range in actual ERO submillimeter emission. Thus, While it is clear that the W02 sample contains a number of  submillimeter-bright EROs at bright $K$-magnitudes it is not reasonable to extrapolate this flux to the entire {\kp}-bright population.

\section{The NIR Color-Color Diagram}

In the above analysis we have treated the entire sample of EROs as a uniform population while the well-known reality is the population is  heterogeneous.  The presence of old elliptical galaxies which have no substantial dust will dilute any submillimeter signal from the dusty galaxies. \citet{poz00} argued that old passively evolving stellar populations may be statistically distinguished from young dusty starbursts based on their locations on a NIR color-color diagram.  The reason for this is straight-forward.  Between redshifts $z\sim$ 1-2 the $RIJK$ filter bands span the 4000A break seen in elliptical galaxies resulting in  (extremely) red $(I-K)$ and $(R-K)$ colors while the $(J-K)$ colors remain moderate.  Dusty starbursts have a relatively smooth SED in this region producing red colors in all filter combinations.   The success of this technique to isolate dusty EROs, at least in a statistical sense, has also been reported by W02. 

\begin{figure}
\epsscale{1.0}\plotone{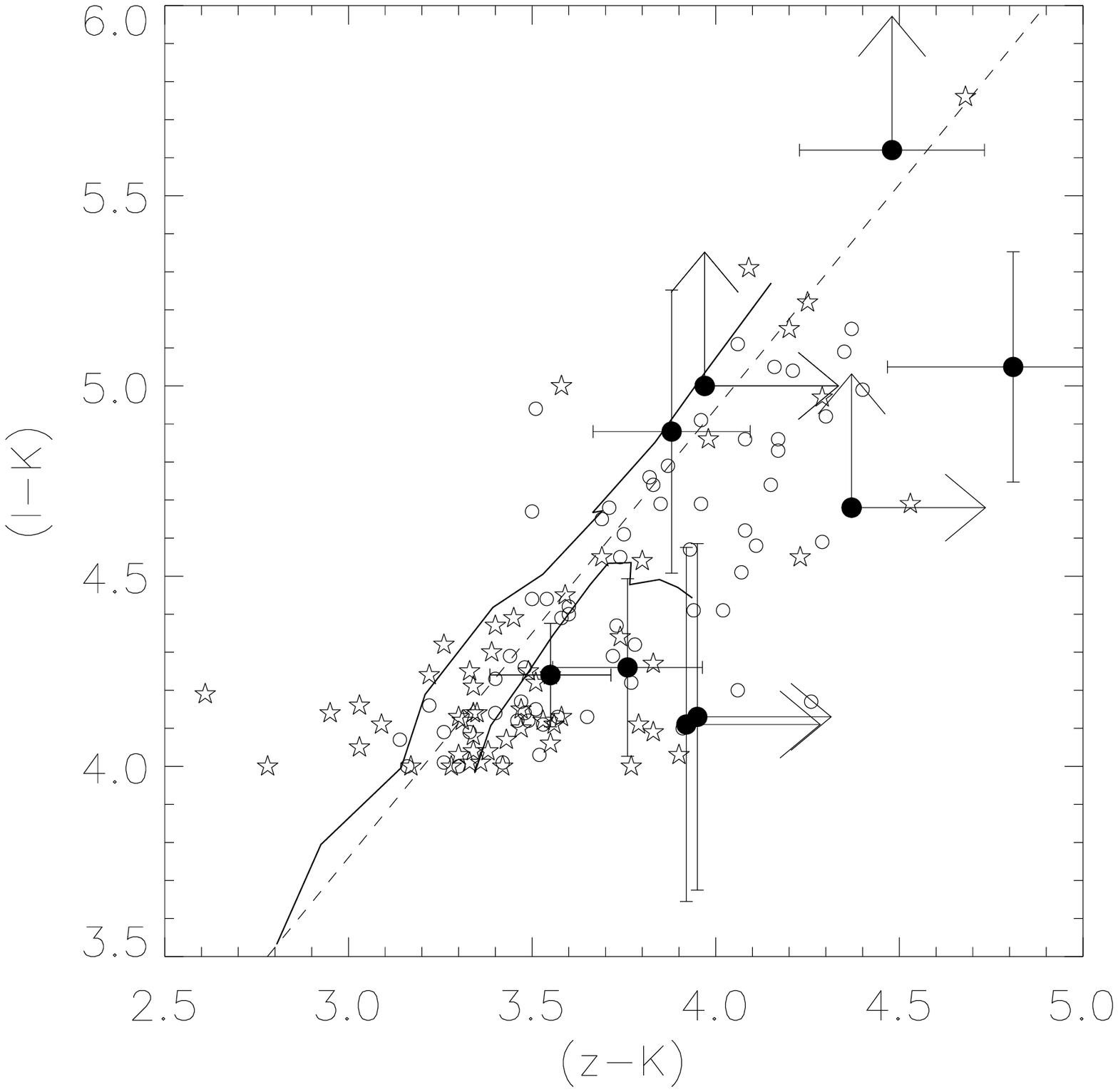}
\epsscale{1.0}\plotone{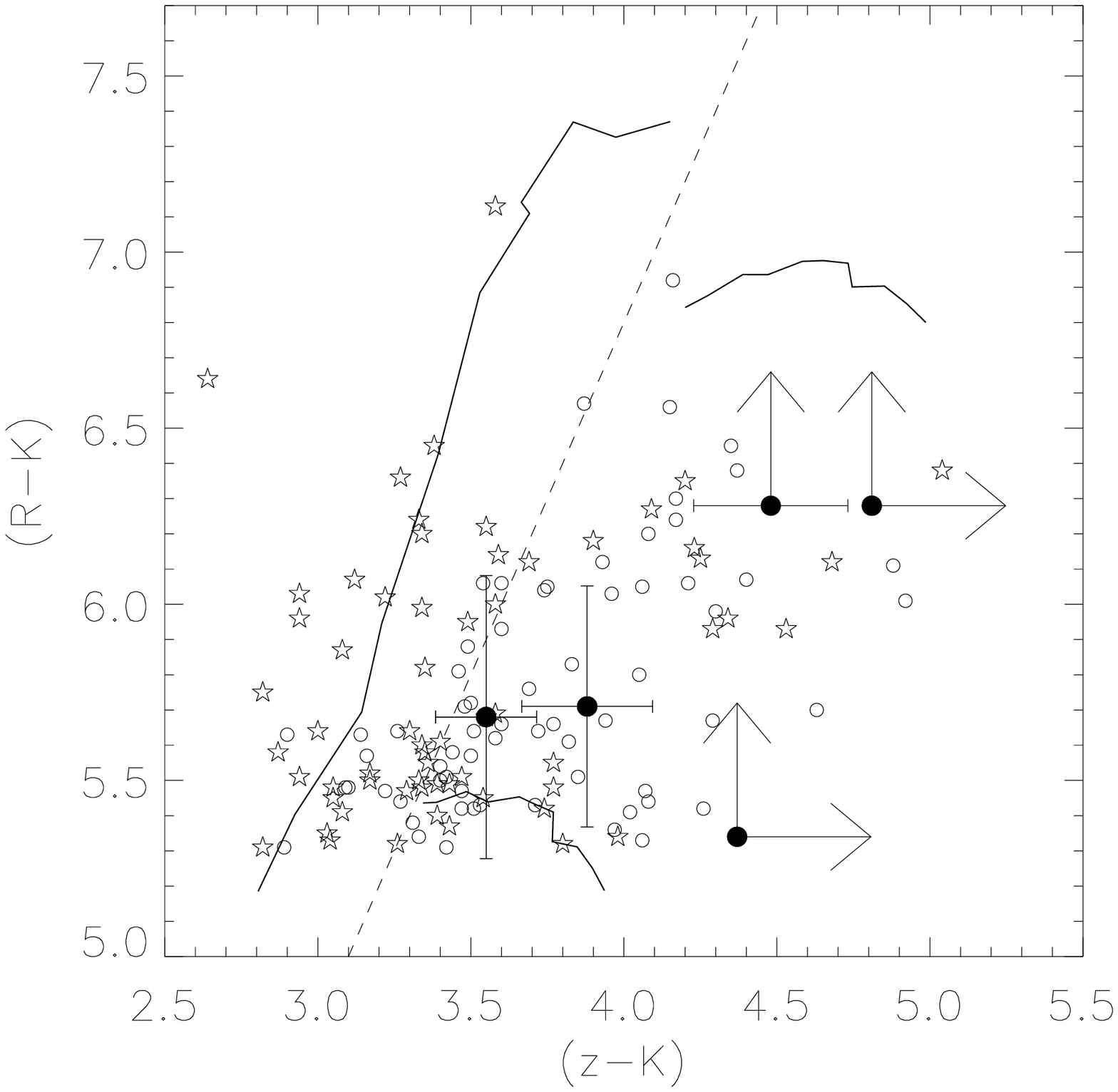}

\caption{ The NIR color-color plots for both ERO selection criteria.  The top figure  shows the $(I-K)$-selected sample and the bottom figure  shows the $(R-K)$-selected sample. Plotted on each are the EROs from both fields.  The open stars denote CUDSS03 objects and the open circles denote CUDSS14 objects.  Strong submillimeter emitters (that is, the most likely counterparts of  $>$ 3mJy sources) are shown with the solid points. In these plots we have only included EROs which have $>$2$\sigma$ detections in $RIz$. The solid lines show the location of the SEDs discussed in the text: the  elliptical shown by the left-most line  and  the dusty starburst  by the right-most line, using the stellar population synthesis code of \citet{bru93}.  Both were formed at $z$= 6 and allowed to evolve.  The dusty starburst has been reddened using the \citet{cal00} law and $E(B-V)$ = 0.8 (for the $(I-K)$ plot) and $E(B-V)$=0.5, 0.8 for the $(R-K)$ plot.    The dotted line corresponds to the approximate dividing line between these two spectral types.    \label{colbc}}
\end{figure}

In Fig.~\ref{colbc}  we show the NIR color-color diagrams for the ERO samples, using $z$ (9000A) in place of $J$.   Also shown are the locations or limits of the EROs identified with discrete submillimeter sources  ($S_{850\micron} >$ 3mJy).  Overlaid are the tracks of two template galaxies ranging from $z$ = 1 - 2: a starburst with dust and an early-type. The dotted line shows the rough division line between areas inhabited by each SED.

This rough division can be used to  group the EROs into those with elliptical-like colors and those consistent with starbursts  provided the redshifts are restricted to $z$ = 1 - 2. Beyond this redshift old early-types will begin to contaminate the starburst side of the plot.   Though the error-bars are substantial the submillimeter luminous objects clearly  lie on the expected star-burst side of the diagram. Ideally one would now repeat  the mean 850$\mu$m flux measurement on each separate group to see if the dusty starbursts as a population  indeed produce more submillimeter emission than the early-types. In this case however, the  original measurement on the heterogeneous population was entirely dominated  by the discrete submillimeter sources which now almost exclusively lie on the starburst side of the plot. Thus, a repeat of this analysis would increase the submillimeter signal (by statistically removing old elliptical-like EROs) but would not provide any new information since we would still only be detecting these bright systems.

\section{ERO companions}

It is well established that the  ERO population is strongly clustered \citep{dad00,roc02,fir02,roc03} over a wide range of magnitudes and color selection criteria,  and this is easily  interpreted in the framework of hierarchical models:  galaxies which are old at $z\sim$ 1 (as a significant fraction of the EROs are) must have formed in the highest density regions in the early universe, and these regions themselves are highly clustered.  Since  separating the old elliptical-like galaxies from the starbursts is not trivial the contribution of each of these groups to the clustering signal difficult is to disentangle. \citet{dad02} however, have shown that  while old ellipticals are indeed highly clustered, the dusty starburst population is not (\citet{dad02}, $K\lesssim$ 19, $(R-K)>$ 5).   \citet{roc03} have investigated the frequency of close pairs of EROs ((I$_{775}-K_s$) $>$ 3.92, $Ks<$ 22) and found  no evidence that dusty star-forming EROs are more likely to exist in close pairs than old passive EROs, or even  optically selected galaxies \citep{pat02}.  Thus, there is very little evidence for strong spatial correlations between dusty starburst EROs.  

The small areas observed thus far with SCUBA have not allowed a reliable clustering measurement for the SMG  population \citep{sco02,web03b,bor03}.  These systems may be  the result of purely stochastic processes and  therefore would not be  clustered, as seen for the local FIR-selected population \citep{mor94,kim98}. An alternative scenario  is one in which the SMGs are strongly clustered, and represent sign-posts for high-density, active  environments at high-redshift.  This latter picture is generally supported by the large  mass estimates for a small number of these systems ($>$10$^{11}$ M$_{\odot}$) \citep{fra99,ivi00,gen03,ner03}. Thus, though dusty EROs as a group may not be highly clustered,  perhaps there is a subset of systems  which are found  in high-density regions and it is these which  are related to  the bright SMG population. If so, this implies environmental effects are driving the star-formation in these systems.

\begin{figure}

\epsscale{1.0}\plottwo{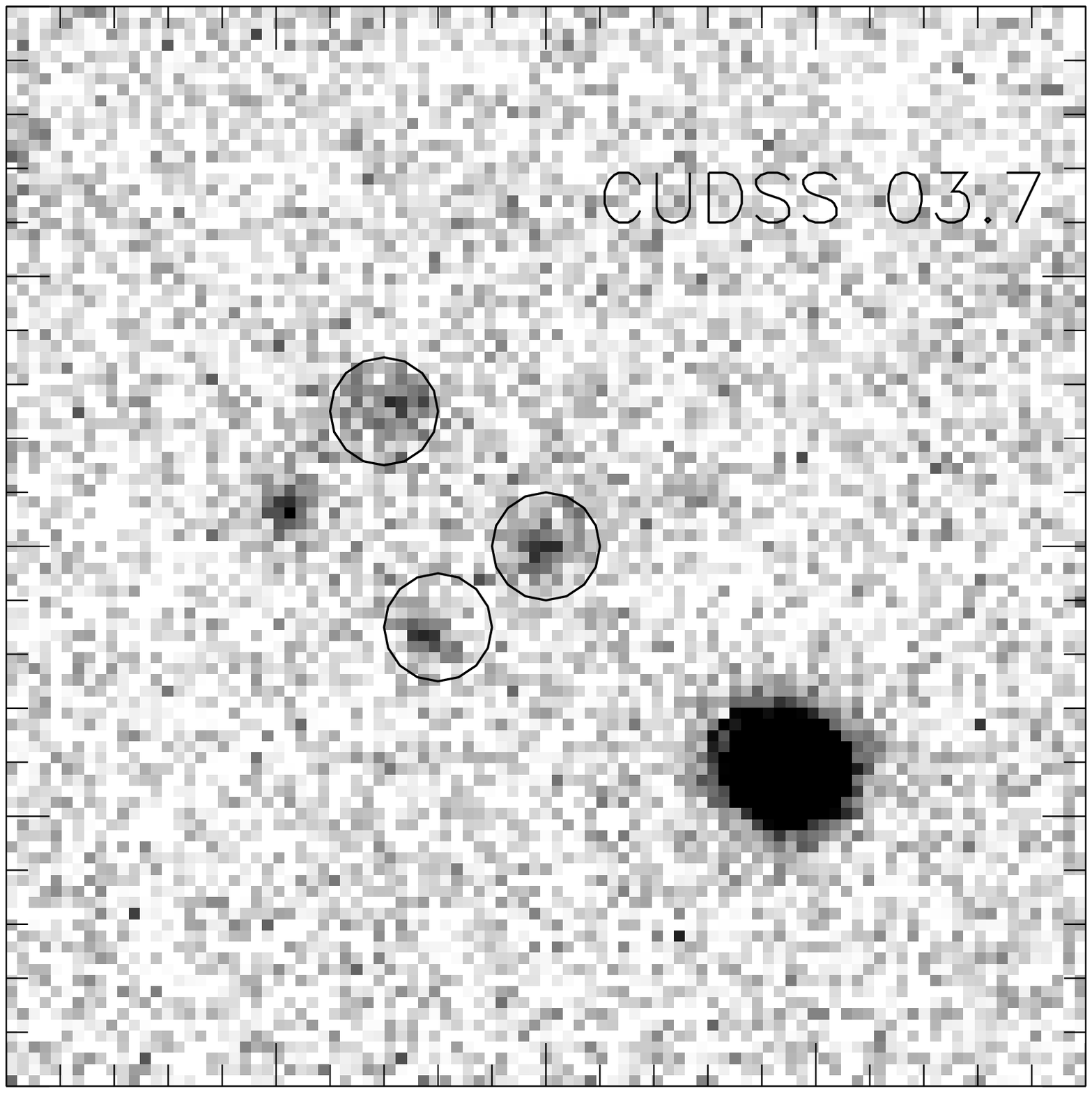}{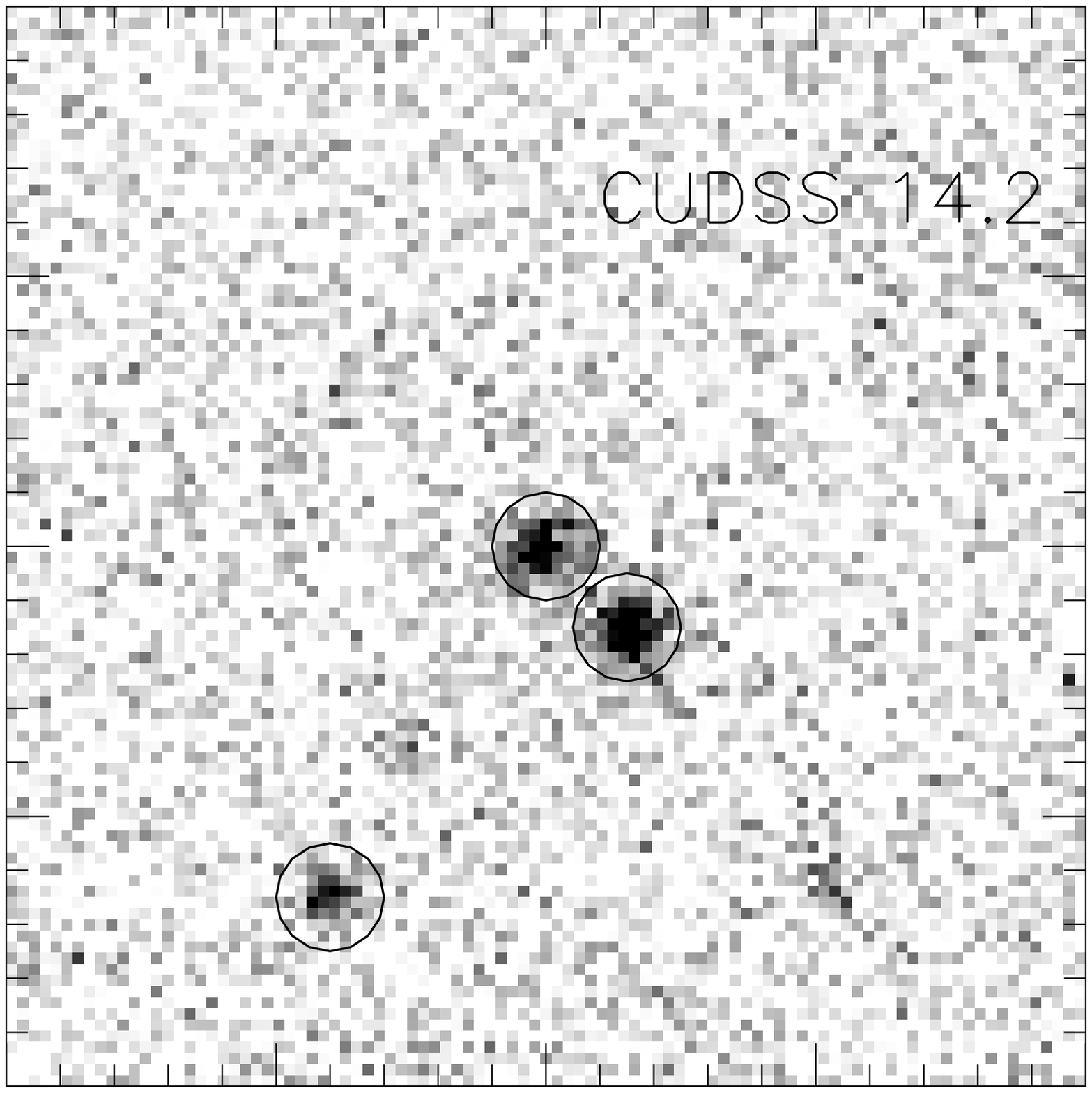}
\epsscale{1.0}\plottwo{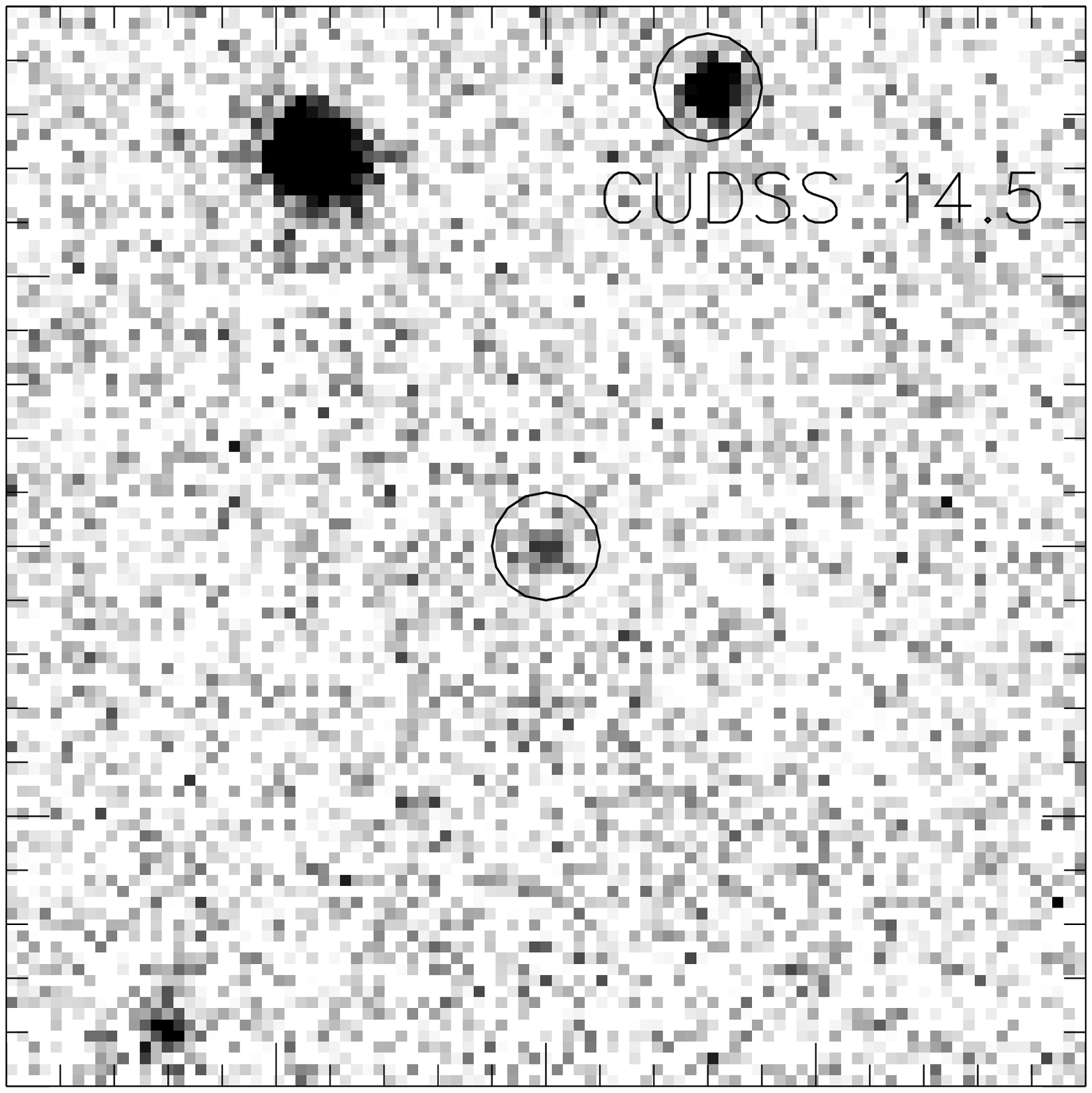}{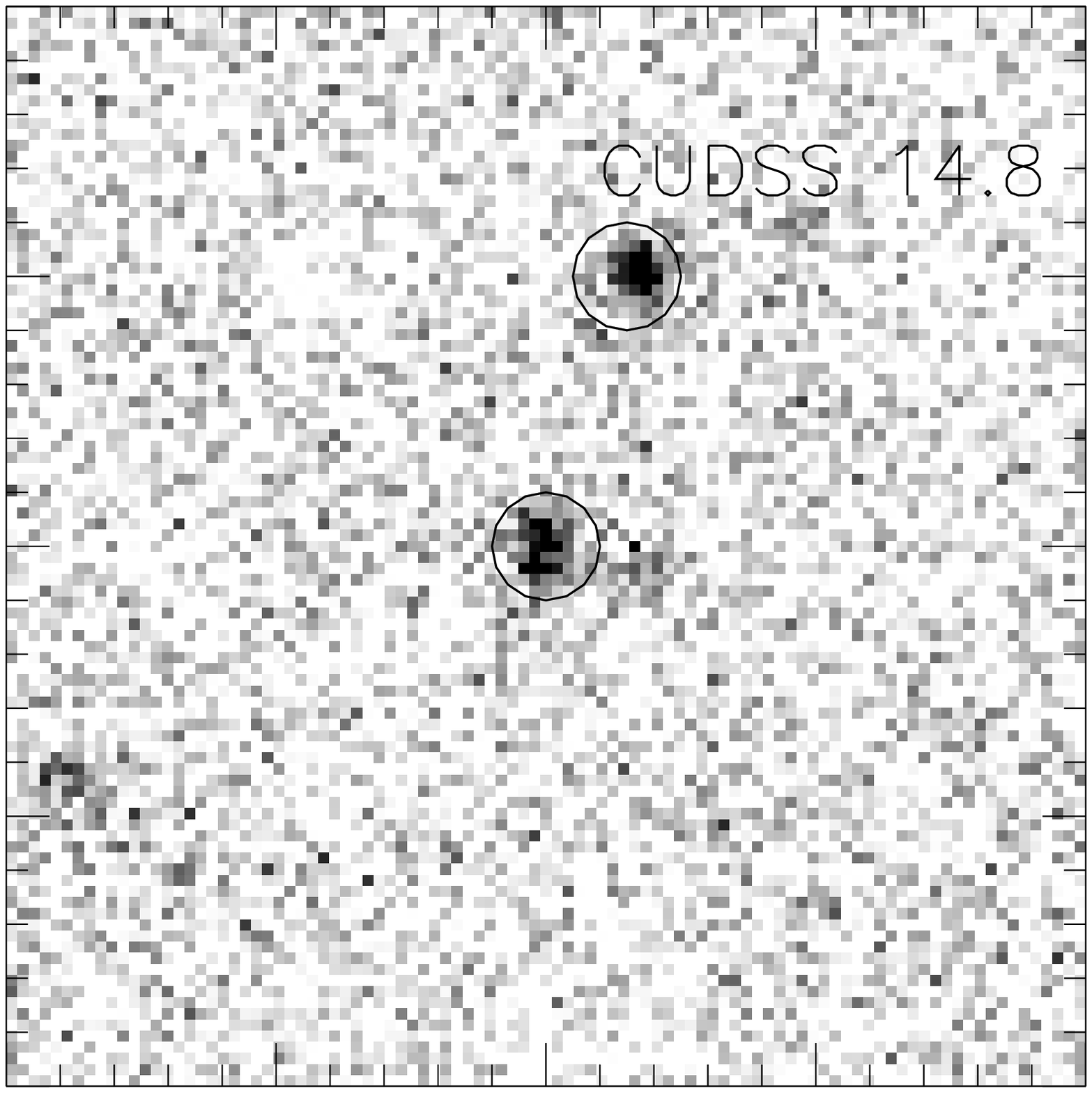}
\centering\epsscale{0.5}\plotone{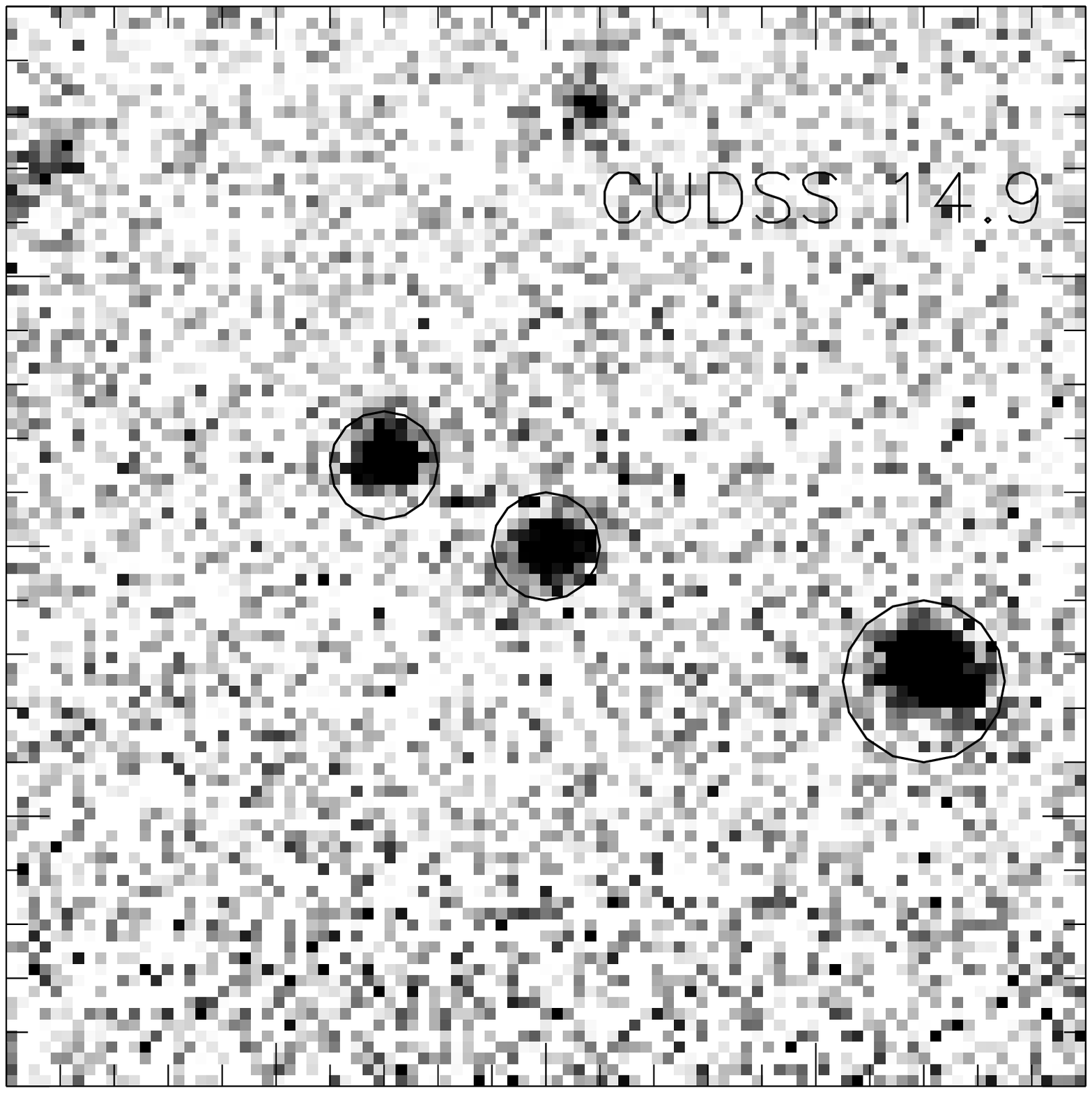}

\caption{The $K$ images (20\arcsec$\times$20{\arcsec})  of the five bright submillimeter sources identified with EROs that have apparent companions.  Each image is centered on the ERO selected as the identification (from statistical analysis) and EROs are marked by circles.  North is up and East is left. \label{postages}}
\end{figure}

There are ten SMGs in the CUDSS catalog which are best identified with EROs (see \S 4.1). Of these, five have another ERO within 10{\arcsec}, and three of these five have two additional EROs within this distance (Fig.~\ref{postages}). To quantify the significance of this we note that a randomly chosen ERO in our sample will have a companion within 10{\arcsec} $\sim$30\% of the time.  As there is no spectroscopic redshift information for these groups of EROs, it is impossible to know if all of these are true physical pairs  or simply projections,  and to determine their true separations.  Nor can photometric redshifts  constrain the redshifts with enough accuracy  since all of these systems are very faint and the photometric  wavelength coverage is insufficient for reliable redshift estimates.  However, there is circumstantial evidence which suggests that most of these pairs are real.   Four out of the five groups consist of EROs with similar apparent $K$ magnitudes and NIR colors (the exception being CUDSS14.5).  If the groups were due to the chance alignment of galaxies at different redshifts such uniformity would not be expected.   This argument is clouded by the fact that one of the EROs in each of the  CUDSS03.7 and CUDSS14.9 groups have radio emission, while their neighbors do not, which could indicate  different redshifts.

The offsets between these EROs range from $\sim$ 2{\arcsec} to 10{\arcsec} which correspond to projected distances of roughly 20-100 $h^{-1}$Kpc at  redshifts of $z\sim$ 1-2.  At these distances the systems are unlikely to be strongly interacting (for example, in the late stages of a merger), but could be bound systems either influencing each other, or experiencing the same global environmental influences.  Due to the faintness of the objects and the quality of the $K$ data the  morphologies are inconclusive: though some objects show asymmetries, there are no clear indications of merger-like activity. 

\begin{figure}
\epsscale{1.0}
\plotone{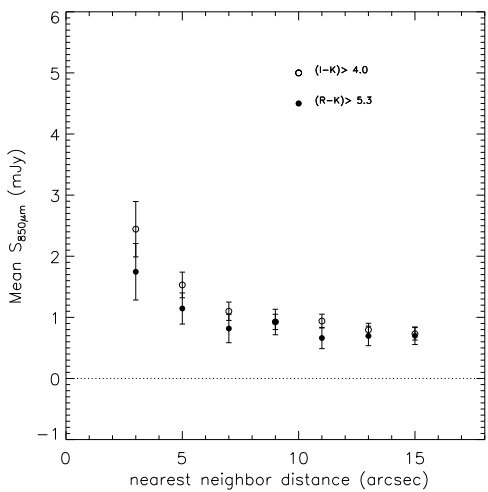}
\caption{In this plot we show the mean 850{\micron} flux of EROs which have at least one other ERO within a given distance (the companion distance).    \label{comp}}
\end{figure}

In Fig.  \ref{comp} we show the mean 850{\micron} flux of EROs as a function of their nearest neighbor distance.  That is we measure the mean submillimeter flux of those  EROs with at least one companion within a given distance, for varying distances.  We find  EROs with  companions  mark on average regions of strong submillimeter emission.  Though these  measurements are dominated by the very bright groups discussed above and shown in Fig \ref{postages} there are very few  close pairs of EROs which do not show evidence of any submillimeter emission.

This result can be interpreted in two ways. Firstly,  the submillimeter emission is produced or dominated by only one ERO within these groups, and this object serves as a marker for an over-density of EROs. This view is supported by radio detection of one ERO in  two  of these systems (CUDSS14.9 and CUDSS03.7) which  likely marks the true location of the submillimeter emission. The isolation of the submillimeter emission to one component of a composite system has been seen by other authors (e.g. \citet{ivi01,ner03}) though generally on smaller angular scales than seen here.  Secondly, all of the EROs confused within the SCUBA beam may be  responsible for the submillimeter emission each at a relatively equal level. Though individually the objects could likely lie below the 3$\sigma$ detection limit of these submillimeter maps,  the combined submillimeter emission from all would result in a single observed  submillimeter source.  This type of effect was clearly seen  during the Monte Carlo noise and source detection analysis of our maps  in  \citet{eal00}.   
 These two scenarios have different implications for the strength of the submillimeter emission from individual objects, but both imply some level of interaction is driving up the average star-formation rate in these systems.

The spatial association between bright SMGs and regions of galaxy over-densities has been previously observed for a number of systems identified with high star formation rates or  AGN activity, including  Ly$\alpha$ emitters, radio galaxies,  EROs, and x-ray sources \citep{ivi00,sma03,ale03}, and indeed between these systems themselves regardless of submillimeter information.    In hierarchical formation models  such regions, characterized by enhanced numbers of extreme objects, may subsequently evolve into the most massive galaxy clusters in the present-day universe.  

\section{The nature of the submillimeter-bright EROs}

Ultimately, we wish to understand why some EROs are so luminous in the submillimeter while others show very little submillimeter emission, if at all.  Obviously, the most dominant and trivial effect is whether an ERO is a dusty star-burst galaxy, rather than an old early-type, but even within the star-bursts there is clear variation.   We have seen in the previous section that environment  plays a role, such that EROs within small groups or pairs often show strong submillimeter emission. 

The strong increase in the mean submillimeter flux of EROs with NIR color seen in Fig.~\ref{colsmmp} could be indicative of a relationship between submillimeter flux and redshift within the ERO population, since a given starburst SED will become increasingly red when placed at higher redshifts.  Thus, the EROs which show submillimeter emission may lie at predominantly higher redshifts than those that do not.  The median redshift of the submillimeter-selected population of SMGs is $z\sim$2-3 \citep{sma02b,ivi02,cha03b}, and so the lower redshift sources in this distribution may over-lap with the high-redshift tail of this ERO sample.  Of course, redder NIR color may simply be an indication of an increased  level of extinction within these EROs, such that redder EROs are more submillimeter-bright because they are dustier \citep{cha03a}.

\begin{figure}
\epsscale{1.0}
\plotone{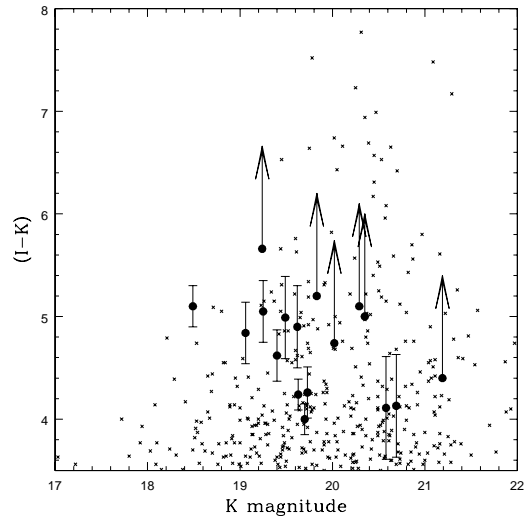}
\caption{A comparison of the NIR properties of the (possibly) submillimeter-bright groups of EROs with the general population of EROs. Solid points denote  EROs which are associated with a discrete SMGs and which are found in pairs or groups of size less than the SCUBA beam.   Crosses correspond to the entire $K$-selected catalog (though only the red objects are shown). Some of these SMG-EROs may not be true submillimeter emitters but may be due to chance alignments. Though very tentative, many of  these EROs are brighter than the background ERO population, for a given color.   \label{colmag}}
\end{figure}

In Fig.~\ref{colmag} we show the color and magnitude of EROs associated with the discrete submillimeter sources. Since we plot all of the EROs within the beam of a discrete source it is possible that   not all of these systems are true submillimeter emitters.  The bulk of the objects with measured colors  are among the brightest objects in the field at each given color.  Again, if we treat NIR color as a very rough redshift indicator then  the submillimeter-bright EROs are  intrinsically more luminous than  average $K$-selected EROs a given redshift. Thus not only are these systems currently experiencing higher than average star-formation rates but they may also have substantial populations of evolved stars compared to the rest of the starbursting EROs (provided they lie at $z\lessim$ 2).   The fact that at least some of these systems exist in over-dense regions  may indicate that  these EROs have experienced a continual series of mergers (even the ones that currently appear isolated)  over an extended time-scale, building up significant evolved populations through a series so strong bursts.

\section{Conclusions}

We investigated the submillimeter properties of EROs in the CUDSS fields and found the following:

\begin{enumerate}
\item{ We detect significant submillimeter flux from EROs in the CUDSS fields of  0.40 $\pm$ 0.07 mJy or 0.56 $\pm$ 0.09 mJy for $(I-K)>$ 4.0 and $(R-K)>$ 5.3 ERO selection criteria, respectively.  This implies EROs to a depth of $K<$ 20.7 produce 7-11\% of the extragalactic background light at 850\micron. However, the flux  is dominated by a minority  of systems which are very bright at submillimeter wavelengths and overlap with the discrete population of submillimeter-bright galaxies discovered with SCUBA.} 
\item{EROs with  strong submillimeter flux  lie on the expected starburst side of the  NIR color-color diagram, as suggested by \citet{poz00}. }
\item{Pairs or small groups of EROs mark, on average, regions of strong submillimeter flux, indicating that star formation rates are enhanced within these high-density areas.  This implies, some fraction of strong submillimeter emitters trace over-dense regions in the early universe.  We are unable to tell from these data if more than one object within these groups is responsible for the submillimeter emission}.    
\item{We tentatively suggest that EROs which show submillimeter emission  are the higher-redshift and higher-luminosity tail of the dusty starburst EROs within these data,  and may have a history of significant and extended  merger activity.}
\end{enumerate}

{\it ACKNOWLEDGMENTS}

We thank the anonymous referee for many helpful comments.
Research by Tracy Webb is supported by the NOVA Postdoctoral Fellowship program.  Research by Steve Eales is funded through the Particle Physics and Astronomy Research Council and Leverhulme Trust.  We wish to thank Natascha F\"orster Schreiber, Doug Johnstone, and Kirsten Kraiberg Knudsen for very helpful discussions.  The JCMT is operated by the Joint Astronomy Center on behalf of the  Particle Physics and Astronomy Research Council in the UK, the National Research Council of Canada and the Netherlands Organization for Scientific Research.  The Canada-France-Hawaii Telescope is jointly operated by the National Research Council of Canada, the Centre National de la Recherche Scientifique of France and the University of Hawaii.

\end{document}